\documentstyle[12pt]{article}
\newlength{\extraspace}
\newlength{\extraspaces}
\newcommand{\bq}{\begin{eqnarray}
\addtolength{\abovedisplayskip}{\extraspaces}
\addtolength{\belowdisplayskip}{\extraspaces}
\addtolength{\abovedisplayshortskip}{\extraspace}
\addtolength{\belowdisplayshortskip}{\extraspace}}
\newcommand{\eq}{\end{eqnarray}}

\newcommand{\newsection}[1]
{\vspace{1mm}
\pagebreak[3]
\addtocounter{section}{1}
\setcounter{equation}{0}
\setcounter{subsection}{0}
\setcounter{footnote}{0}

\begin{flushleft}
{\large\bf \thesection. #1}
\end{flushleft}
\nopagebreak
\medskip
\nopagebreak}

\newcommand{\newsubsection}[1]{
 \vspace{0.5mm}
\pagebreak[3]
\addtocounter{subsection}{1}
\noindent{ \bf \thesubsection. #1}
\nopagebreak
\vspace{1mm}
\nopagebreak}

\begin{document}

\addtolength{\baselineskip}{.8mm}

\def\tL{\theta_{L}}
\def\tR{\theta_{R}}
\def\o{\omega}
\def\A{{\cal A}}
\def\ot{{1\over 2}}
\def\lsim{\mathrel{\rlap {\raise.5ex\hbox{$ < $}}
{\lower.5ex\hbox{$\sim$}}}}
\newcommand{\vsone}{\vspace{1cm}}
\newcommand{\pr}{\paragraph{}}
\newcommand{\be}{\begin{equation}}
\newcommand{\ee}{\end{equation}}
\newcommand{\bea}{\begin{eqnarray}}
\newcommand{\nn}{\nonumber}
\newcommand{\eea}{\end{eqnarray}}
\newcommand{\nd}[1]{/\hspace{-0.6em} #1}
\newcommand{\nk}{\noindent}
\baselineskip=18pt

\def\gappeq{\mathrel{\rlap {\raise.5ex\hbox{$>$}}
{\lower.5ex\hbox{$\sim$}}}}

\def\lappeq{\mathrel{\rlap{\raise.5ex\hbox{$<$}}
{\lower.5ex\hbox{$\sim$}}}}

\thispagestyle{empty}

\begin{flushright}
{\sc OUTP} -95 - 50 P\\
 hep-th/9512210 \\
 December  1995
\end{flushright}
\vspace{.3cm}

\begin{center}
{\large\sc{ World-Sheet Logarithmic Operators and
 Target Space Symmetries   in String Theory   } }\\[15mm].

{\sc  Ian I. Kogan}  and {\sc Nick  E. Mavromatos}
\footnote{P.P.A.R.C. Advanced Fellow}\\[2mm]
{\it Theoretical Physics, 1 Keble Road\\[2mm]
      Oxford, OX1 3NP, UK} \\[15mm]

{\sc Abstract}

\end{center}
 We discuss  the  target-space interpretation of the  world-sheet
logarithmic operators in  string theory.  These operators
 generate  the normalizable zero modes (discrete states)
 in target space, which restore the symmetries of the theory
 broken by the background.  The problem of the recoil in string
 theory is considered, as well as some general properties of
 string amplitudes containing logarithmic operators.

\noindent

\vfill
\newpage
\pagestyle{plain}
\setcounter{page}{1}
\stepcounter{subsection}
\newsection{Introduction.}
\renewcommand{\footnotesize}{\small}

Recently it was discovered  that in some conformal field theories,
  namely  the c = - 2  model
\cite{gurarie}, the gravitationally dressed  CFT models  \cite{bk},  the
$c_{p,1}$ models with central charge $c = 1 - 6(p-1)^2/p$  \cite{flohr}
and some critical disordered models \cite{tsvelik},  there are
 new objects - the so called logarithmic operators,
associated with non-standard
properties of operator products in
the model. For information, we mention
that the logarithmic
 operators in the $c=-2$ model were  discussed  recently
in
some detail in ref. \cite{kausch}.
The existence of logarithmic operators
was first shown by Gurarie in \cite{gurarie}.
 It was pointed out
  that the appearance of logarithms in the $c=-2$-model's
  correlation  functions
 (which also emerged earlier  in the WZNW model
 on the supergroup $GL(1,1)$  \cite{rs}) is
  due to the presence of special operators, whose operator product
expansions (OPE's) display
logarithmic short-distance singularities. These logarithmic
operators have conformal dimensions degenerate with those of the
usual primary operators, and it is this degeneracy that is at the
origin of the logarithms.  As a result of this
degeneracy one can
no longer completely diagonalize the Virasoro operator $L_0$, and the
new operators together with the standard ones form the basis of the
Jordan-cell for $L_0$.  In the paper \cite{tsvelik} it was shown
 that the anomalous dimensions of logarithmic operators must be integers,
 and it was conjectured
 that they  are connected with some hidden
continuous symmetry (which may be a replica symmetry in case of
 the disorder).

In this paper we would like to address the issue of logarithmic
 operators in string theory and to study
the connection between  world-sheet logarithmic operators
 and target-space symmetries.  It turns out that
  these operators are ultimately
  connected with the existence of  normalizable zero modes
 for some backgrounds in target-space. These modes  exist
  when the background violates some symmetries of the theory,
 in which case there is a family of backgrounds connected by
  the  symmetry (or symmetries) which acts on the
  moduli space of parameters characterising the background.
   The simplest example of
  the normalizable zero modes are the zero modes for solitons or
 instantons arising due to translations and rotations (both in the
 space-time and in the internal space) of the backgrounds.

  These modes are crucial
 in calculating the recoil effects in scattering  off the soliton,
 when the soliton state changes  during the process of the scattering.
  Recently this problem was considered in an interesting paper
 \cite{paban} where the effects of recoil in string theory,
 or in  other words the quantization of the collective coordinates
 in string theory,  were discussed  and
  connected with
  the existence of  non-local  world-sheet operators violating
 the conformal invariance and leading to some unexpected logarithms.

 We shall demonstrate
in this article
that these operators are nothing but the
 logarithmic operators mentioned above.
 This statement is true in general,
and is not
restricted to the models considered in \cite{paban}.
This
 makes the world-sheet logarithmic operators very important objects
 in string theory.  There is an important difference between the
 approaches  suggested in ref. \cite{paban} and ours - in the first case
 the logarithmic singularities come from the infinite
  normalization factor
 of some world-sheet current, which is assumed to obey
normal operator product expansion relations for currents.
This is a result which we do not agree with.  In our case,
the extra logarithmic dependences in correlation functions
of the string in the (extended-object) background
 arise simply by identifying these currents with
 logarithmic operators in the theory, which  have well-defined
  Zamolodchikov metric \cite{zam}.

 Another example of a model where these operators are  important
 is the 2d black holes.
 We shall argue, in this case,
 that the target-space symmetry connected with these operators is
 the $W_{\infty}$ acting on the discrete states \cite{emn} in the
 black hole background. It should be noted, in this context, that
extra world-sheet logarithms have appeared in the amplitudes of
non-critical strings propagating in black-hole backgrounds,
which have been interpreted~\cite{emn} as expressing back-reaction
of matter on the space-time geometry, and, therefore, a change
of conformal field theory (CFT). Let us note that, in general,
a change of CFT can be attributed to the Liouville field,
whose connection
with the world-sheet Renormalization-Group flow and evolution process -
the latter
being viewed as a change of CFT backgrounds that may cause
quantum mixing - was discussed in refs. \cite{mixing}.
A detailed connection of these results with the logarithmic operators
discussed in this article will be discussed elsewhere.

 In this work,
 we shall argue about
the emergence of
 logarithmic operators in these models
in two ways: first,  by using
 the degeneracy between  a vertex operator
 corresponding to the  normalizable zero mode and the identity
 operator; second, by demonstrating that the
 conformal blocks in the corresponding
Conformal Field Theory (CFT) possess logarithmic terms.
 These general arguments will be  based on some  specific
 properties of string amplitudes
containing world-sheet logarithmic operators, as well as
on the construction
of   the   Zamolodchikov metric \cite{zam} pertaining
to these operators.
 Before addressing these  problems,
we consider it as useful to
 remind the reader
 some  facts about the logarithmic operators.

\newsection{Logarithmic operators}
\newsubsection{Mathematical background}

 First of all, let us discuss
 how  logarithms appear in the correlation
 functions of  the conformal field theory.
 Let us consider for example
  the four-point correlation functions of
 primary fields $V(z)$ and $V^{-1}(z)$ with anomalous dimensions $h$,
 such that $<V(z) V^{-1}(0)> = z^{-2h}$. This correlation function
 can be represented as
\bea
<V(z_1) V^{-1}(z_2) V^{-1}(z_3) V(z_4)> =
\frac{1}{(z_1 - z_4)^{2h}(z_2 - z_3)^{2h}}  F(x),
\label{F}
\eea
where $x = (z_1 - z_2)(z_3 - z_4)/(z_1 - z_4)(z_3 - z_2)$
 and  only the dependence on holomorphic coordinates $z_i$
 has been written, omitting the obvious
 dependence on  $\bar{z}$. In many different models the
 unknown function $F(x)$ is defined (up to some factors like
$ x^{\alpha}(1-x)^{\beta}$ where $\alpha$ and $\beta$ are some
 parameters depending on the model) as a solution of the
 hypergeometric equation
\bea
x(1-x)\frac{d^2 {\cal F}}{dx^2} +
[c - (a+b+1)x] \frac{d{\cal F}}{dx} - ab {\cal F} = 0
\label{hypereq}
\eea
 which in general has two  independent solutions \cite{ww}, \cite{gr}
 \bea
{\cal F}_1 = F(a,b,c; x), ~~~~~~~
{\cal F}_2 = x^{1-c} F(a-c+1,b-c+1,2-c; x)
 \eea
where $F(a,b,c; x)$ is a hypergeometric function and these two
independent solutions correspond to the  two primary fields in the
 OPE of
\bea
 V(z) V^{-1}(0) = \frac{1}{z^{2h}}~[ I + z^{c-1} O  + ...]
\label{OPE1}
\eea
one of which is an identity operator $I$
 and the second  operator $O$  (adjoint operator
 in the case of the WZNW model, for example, see \cite{kz}) has an
 anomalous dimension $1-c$. In this way we recover the
 conformal blocks in a generic CFT \cite{bpz}.
 However, this is not true anymore if the parameter
  $c$ in the  hypergeometric equation (\ref{hypereq}) is an
 integer (which takes  place in all examples in \cite{gurarie} -
 \cite{tsvelik}).  There  is a  general  theorem in  the theory
 of  the second order  differential equations which deals with
 the expansion of the solution near the regular point
   $x=0$
\bea
x^{\alpha}\sum_{n} a_n x^n
\label{ryad}
\eea
 This theorem  tells us how to calculate the coefficients $a_n$ if one
 knows  the two roots $\alpha_1$ and $\alpha_2$, which are the
 solutions of  the  so-called indicial equation \cite{ww}.
 However, if the difference $\alpha_1 - \alpha_2$
 is an integer, the  second solution either equals the first one
 (when $\alpha_1 =\alpha_2$) or some of the coefficients are undefined.
  In both cases the second solution  has
  logarithmic terms $x^n \ln x$ in the expansion (\ref{ryad}), besides
the usual terms $x^n$.  For hypergeometric
 equation the indicial equation is
\bea
\alpha(\alpha - 1 + c) = 0
\eea
and the two roots are $\alpha_1 = 0$ and $\alpha_2 = c-1$, i.e.
 for integer $c = 1+ m$ the second solution has logarithmic terms.

Let us note that in this case the second operator $O$ in the OPE
(\ref{OPE1}) is degenerate either with the identity operator
 (when $m=0$ and $c=1$) or  with  one of its Virasoro descendants
 (for negative integer $m$ when $O$ has a positive  dimension
 $|m|$). For positive  integer $m$ ( in which case $O$ itself
 has a negative dimension $-m$) one of its descendants will be
 degenerate with $I$.

The condition that $c= 1+ m,~~m \in Z$ is necessary, but not
sufficient if $m \neq 0$. To have logarithms in this case
 one has to impose additional
 constraints on $a$ and $b$ \cite{ww},\cite{gr}.

 If $c= 1 + m$, where $m$ is a natural number,
  the two independent solutions  are
\bea
{\cal F}_1 = F(a,b,1+m; x), \nonumber \\
{\cal F}_2 = \ln x~ F(a,b,1+m; x) + H(x),
\label{1+m}
 \eea
where $H(x) = x^{-m}\sum_{k=0}^{\infty}h_k x^k$ and $h_{m} =0$,
unless either $a$ or $b$ equal $1+m'$ with $m'$ a natural number
 $m'<m$. In this case  the second solution is only a polynomial in
$x^{-1}$ and the are no logarithms at all.

 If $c= 1 - m$, where $m$ is a natural number,
  the two independent solutions  are
\bea
{\cal F}_1 = x^m F(a+m,b+m,1+m; x), \nonumber \\
{\cal F}_2 = \ln x~ x^m  F(a+m,b+m,1+m; x) + H(x),
\label{1-m}
 \eea
where $H(x)$ is again some regular expansion without logarithms,
unless either $a$ or $b$ equal $-m'$ with an integer $m'$
such that  $0 \leq m'<m$, in  which  case  both solutions do
not have  logarithmic terms also.

\newsubsection{Amplitudes in string theory with logarithmic operators}

  The logarithmic terms in the conformal blocks can not be explained
 by usual the OPE and it is necessary to introduce new operators
 in the theory - the logarithmic operators. The OPE for
 primary fields $V$ and $V^{-1}$  takes the form
\begin{equation}
V(z_1)~ V^{-1}(z_2) = ... + (z_1-z_2)^{h_C - 2h }
(\bar{z}_1-\bar{z}_2)^{\bar{h}_C - 2\bar{h} }
\left[D + C \ln|z_1-z_2|^2 \right] + .... ,
\label{AB}
\end{equation}
where the dimension $h_C$ of the operators $C$ and $D$ is determined by
the leading logarithmic terms in the conformal block (\ref{F}).
  We have
  written here both $z-$ and  $\bar{z}-$dependences explicitly
 and it is important to note that the logarithmic term depends on
$|z|$, even for chiral fields, because in the full conformal blocks
 actually $\ln |z|$ appears, as was shown in \cite{tsvelik}.

 Some general  properties of these operators
were studied in \cite{gurarie} and
 \cite{tsvelik}.   The simplest example of a logarithmic operator
 one can consider
  is  the  puncture operator, which
  was discussed in \cite{polch} in the context of the
 Liouville  model with the action
\begin{equation}
S = \frac{1}{8\pi} \int \mbox{d}^2\xi \sqrt{g(\xi)}\left[
\partial_{\mu}\phi(\xi) \partial^{\mu}\phi(\xi) + Q R^{(2)}(\xi)
\phi(\xi)
 \right]
\label{Liou}
\end{equation}
 where $R^{(2)}$ is a world-sheet  curvature.
 The  central charge equals
\begin{equation}
c_{L} = 1 + 3 Q^2 \label{central}
\end{equation}
 and the primary fields
 $\exp(\alpha \phi)$ have  dimensions
\begin{equation}
h_{\alpha} =
 \alpha(Q-\alpha)/2 \label{Eq}
\end{equation}
 This means that there are two cosmological constant
 operators with the same dimension $h_{\alpha}=1$, namely
 $V_{\pm} = \exp( \alpha_{\pm} \phi)$, where
$$
\alpha_{\pm} = \frac{Q}{2} \pm \frac{1}{2}\sqrt{Q^2 -
8}
$$
If  $Q^2 = 8$, i.e.  $c_L = 25$,
   there is a degeneracy $\alpha_{+} =
 \alpha_{-} = \sqrt{2}$ and instead of two exponential primary fields we have
 only  one  $C = \exp(\sqrt{2}\phi)$. The  second field with the same
 dimension (which is called the puncture operator) turns out  to be
 $ D = (1/\sqrt{2})\phi \exp(\sqrt{2}\phi)$,
 here $(1/\sqrt{2})$ is the normalization factor.

 It is easy to show that  the OPE of the stress-energy
 tensor $T$ with these fields is   the follows
\begin{eqnarray}
T(z) C(0) &=&  {h \over z^2} C(0)+ {1\over z} \partial_z C(0) +
...
\nonumber\\
T(z)  D(0) &=&  {h \over z^2} D(0)+{1\over z^2}
 C(0)+{1\over z} \partial_z D(0) + ...
\label{JOPE}
\end{eqnarray}
 which is written here for  general logarithmic pair
   $C$ and $D$ with anomalous dimensions $h$.
 This OPE obviously leads to  a  mixing  between $C$ and $D$.
 The  Virasoro operator $L_0$  which is defined through the
 Laurent expansion $T(z) = \sum_{n} L_{n} z^{-n-2}$ is not diagonal
 and mixes   $C$ and $D$
inside a $2\times 2$ Jordan cell
\be
L_0 |C> =h |C>; \qquad L_0|D>=h |D> + |C>
\label{oneb}
\ee

The OPE (\ref{JOPE}) leads to a modification of the  Ward identity for the
 string amplitudes
\bea
\int d^2 z_1...d^2 z_N~ <T(z) V_{1}(z_1)....V_{N}(z_N)>
\eea
where $V_{i}(z_i)$  are primary fields with dimensions $h_i$.
In the usual case  one can show
\cite{polyakov}
that for on-shell states,i.e. when  dimensions $h_i = 1$,
 these amplitudes are zero, which means that the descendants
 $L_{-n_1}..L_{-n_k} V_{i}$ decouple from the physical amplitudes.
 This is not true in case when some of the fields $V_i$ are
logarithmic operators. In this case, using the OPE (\ref{JOPE})
 one can see that (all $h_i =1$)
\bea
\int d^2 z_1...d^2 z_N~ <T(z) V_{1}(z_1)...D_j(z_j)...V_{N}(z_N)>
\nonumber \\
= \int d^2 z_1...d^2 z_N~ \sum_i  {\partial \over \partial z_{i}}
\left( {1\over (z-z_i)}
< V_{1}(z_1)...D_j(z_j)...V_{N}(z_N)> \right)
 \label{Wi}
 \\
 + \int d^2 z_1...d^2 z_N~ \sum_{j}{1\over (z-z_j)^2}
< V_{1}(z_1)...C_j(z_j)...V_{N}(z_N)> \nonumber
\eea
The first sum is a total derivative and after integration gives zero,
 but the second one, where the summation is performed only on $j$
 corresponding to the $D$ operators, is non-zero and this Ward
identity connects  amplitudes of the descendants of $D$ with
 with amplitudes without  logarithmic operators (but including
 currents $C$). One can derive a Ward  identity for $\bar{T}$,
 which also will be nontrivial. Several examples of these identities
 were discussed in  \cite{tsvelik}.

 The important lesson we learn from this is that both $T$ and $\bar{T}$
 descendants of logarithmic operators do not decouple from physical
 amplitudes - this will be important below. Some examples of this
type of Ward identities
 were considered in \cite{tsvelik}.

 Substituting  the OPE (\ref{AB}) into  the four-point correlation
function one can derive (see details in \cite{tsvelik}) the following
 two-point correlation functions for the fields $C$ and $D$
\begin{eqnarray}
\langle C(x) D(y)\rangle =
\langle C(y) D(x) \rangle  = \frac{\kappa}{2~(x-y)^{2h_C }}
\nonumber \\
\langle D(x) D(y)\rangle =
 \frac{1}{(x-y)^{2\Delta_C}} \left(-\kappa\ln|x-y|^2 + d\right)
 \label{CC}
 \\
\langle C(x) C(y)\rangle  = 0
\nonumber
\end{eqnarray}
Here the constant $d$
can be made arbitrary by shifting $D \rightarrow D
 + $ const $C$. The coefficient $\kappa$ is defined by the leading
 logarithmic term in the conformal block ( $-2c$ in the
 notation  of ref. \cite{tsvelik}).

 This mixing   also affects the string propagator on the
world-sheet cylinder between states $|m>$ and $|n>$,
 which is defined as
\bea
\int dq d{\overline q} <n|q^{L_0 -1}  {\overline q}^
{{\overline L}_0 -1} |m>
\eea
where $q = \exp (2\pi i \tau)$ and $\tau$ is the modular parameter.
In the usual case, when $L_{0},~ {\overline L}_0$ are diagonal,
 one gets after integrating over $\tau$
\bea
 <n|\frac{1}{L_0 +{\overline L}_0 -2}~\delta(L_{0}- {\overline L}_0) |m>,
\eea
where $\delta(L_{0}- {\overline L}_0)$ enforce the condition $h =
\bar{h}$ for all propagating states.
 However in the case of logarithmic operators
  one  must  take into account the  Jordan
 cell structure of $L_0,~{\overline L}_0$,
 which in the sector of $|CD>$ and
$|{\overline CD}>$
 states leads to
\bea
q^{L_0} = q^{h_C}
\left(
\begin{array}{cc}
1 & \ln q \\
 0 & 1
\end{array}\right); ~~~~
q^{{\overline L}_0} = {\overline q}^{\bar{h}_C}
\left(
\begin{array}{cc}
1 & \ln {\overline q} \\
 0 & 1
\end{array}\right)
\eea
 We have the new  logarithmic factors here which will be very
 important soon.  Let us also note that
 $\ln q$ factors  arise also in  the characters $Tr_{h}q^{L_0}$
  as was  discussed by Flohr in \cite{flohr}.

 It is important to consider  two separate cases. The first one
is when
 states are  the left-right logarithmic states
$(C,D)(\bar{C},\bar{D})$,
 then   the propagator takes the form
\bea
\int dq d{\overline q}q^{h_C -1}  {\overline q}^
{\bar{h}_C -1} <CD|\left(
\begin{array}{cc}
1 & \ln q \\
 0 & 1
\end{array}\right)|CD><{\overline C}{\overline D}|\left(
\begin{array}{cc}
1 & \ln \bar{q} \\
 0 & 1
\end{array}\right)|{\overline C}{\overline D}>
\label{CDprop}
\eea
One has either $\ln q$ or $\ln \bar{q}$ terms for transitions when either
$D$ goes to $C$ or $\bar{D}$ goes to $\bar{C}$ and $\ln q \ln \bar{q}
 = 4\pi^2 |\tau|^2$ for transition $D\bar{D}$ to $C\bar{C}$.
 This type of states was considered in \cite{tsvelik}.

There is another type of states which we shall call chiral logarithmic
states, when the logarithmic operators are present only in left or
 right sectors, like $L_{-1} \bar{D} + \bar{L}_{-1} D$ when $h =1$,
 or more general descendants for other integer  $h$.  It was
 shown in the discussion following
  the Ward identity (\ref{Wi}) that these
 states in general do not decouple from the physical spectrum.
  In the case of the chiral logarithmic states
 only one Jordan cell - either $L_0$ or $\bar{L}_0$ will work
 and we shall have the sum of $\ln q$ and $\ln \bar{q}$, i.e.
\bea
\int dq d{\overline q}q^{h_C -1}  {\overline q}^
{\bar{h}_C -1} \ln |q|^2
\eea
 for transition between $D$ and $C$. As we shall see later
  these  states will be the central issue of the recoil problem,
 which we shall address now.

\newsection{Logarithmic operators, background symmetries and
 the recoil problem in string theory }
Let us first review the situation in critical string theories
in a given background $\{ g^i \}$, where $g^i$ denotes
generic background couplings/deformations that have vanishing
$\beta ^i (g)$ functions. To be more specific
one considers the following conformal field theory
\be
    S=\int d^2 z  g^i(X) V_i(\partial _\alpha X, {\cal J});
\qquad  \beta ^i(g) \equiv dg^i/d \ln\mu = 0
\label{two}
\ee
where $V_i (\partial _\alpha X, {\cal J}) $ are vertex operators
(deformations) turning on the background $g^i(X)$,
probably depending on some Kac-Moody currents of the
{\it exact}
conformal model representing
string propagation in this background,
$\{ X \}$ are target space-time co-ordinates of the string,
and $\partial _\alpha  X$ are world-sheet
derivatives of the `fields $X$'.
As we shall argue below,
the form of the vertex
operators, depending only on world-sheet derivatives of $X$, and
not on $X$ itself,
will be quite crucial for the demonstration of the existence
of non-trivial zero modes~\cite{paban}.

Let us concentrate on the soliton background of ref.
\cite{paban} for definiteness, keeping in mind that the same
 can be applied  to any other extended object background.
Our aim is  to identify the deformations
that can cause a change of state of the monopole
background during a scattering event  (`recoil').
As we have said earlier this problem is equivalent
to identifying the relevant deformations of the $\sigma$-model
that cause a change in conformal field theory~\cite{emn}.
The important point to notice is that in the presence
of the soliton  background of ref. \cite{paban}
target space translational invariance is broken  by
the center of mass of the soliton.
This effect can be seen in the $\sigma$-model path-integral
formalism
by performing a constant shift of the {\it spatial}
coordinates $X$ of space-time
\be
         X^{\mu}(z, \bar{z})
 \rightarrow X^{\mu}(z, \bar{z}) + q^{\mu}; \qquad
 q^{\mu}=const
\label{three}
\ee
By  expanding $g^(X + q)$ in Taylor series  and taking into account
the specific dependence of $V_i (\partial _\alpha X, {\cal J})$
on world-sheet derivatives of $X$ only,
one observes that the effect of the translation (\ref{three})
is simply to induce a {\it deformation} in the $\sigma$-model
action of the form
\be
  \delta S = q ~ \int d^2 z
 ~\frac{\delta  g^i}{\delta X} V_i\left(\partial _\alpha
X, {\cal J}\right)
\label{four}
\ee
where appropriate summation over spatial indices  in $X$ is assumed.
Such a deformation can be expressed as a total world-sheet
derivative
by using the equations of motion stemming from the
deformed world-sheet action
\bea
\frac{\delta S}{\delta X(z,\bar{z})}=
\partial _\alpha \left(\frac{\delta S}{\delta
( \partial _\alpha X(z,\bar{z}))} \right)
\eea
 The result is~\cite{paban}
\be
       {\cal O}={\cal N}
       (ghosts) \otimes \partial _\alpha \{ g^i(X)
       \frac{\delta}{\delta( \partial _\alpha X)}
V_i (\partial _\beta X ; {\cal J}) \} \equiv
{\cal N} (ghosts) \otimes \partial _\alpha J^\alpha
\label{five}
\ee
where $J^\alpha $ is a two-dimensional Noether current
associated with the translation (\ref{three}),
and the ghost insertions must be included to absorb the
 ghost zero modes.

The constant ${\cal N}$ is a normalization factor which is found
to be logarithmically divergent~\cite{paban}.
 This  was found by
  considering the (normalized) Zamolodchikov metric
corresponding to ${\cal O}$,
 which is by construction~\cite{zam}
\bea
&~&G_{{\cal O}{\cal O}} =4\pi \delta ^{ij}
=|z|^4 <{\cal O}^i(z,{\bar z}),
{\cal O}^j (0) >= \nn \\
&~&{\cal N}^2  ~
|z|^4 < (\partial {\overline J}(z,{\bar z}) + c.c.)
(\partial {\overline J}(0,0) + c.c.) >
\label{six}
\eea
 If one assumes   that in the soliton  background
of ref. \cite{paban} $J$ are usual $(1,0)$ currents,
 one has
\be
 <J^i(z)J^j(0)> \sim \frac{\kappa \delta ^{ij}}{z^2}
\label{seven}
\ee
where $\kappa $ is a normalization constant, and
the Latin indices $i,j$ run over  the set of  zero-modes pertaining to
the specific background.
Then
one obtains
\be
{\cal N}^2 |z|^4 (\partial ^2 \frac{1}{{\bar z}^2} + c.c.) =4\pi
\label{eight}
\ee
 This equation appears obscure to us, and the way how
the normalization constant was obtained in \cite{paban}
is not mathematically clear. Indeed, the authors
of ref. \cite{paban}
integrate the equation over $z$  with a conformally invariant weight
$\int d^2z/|z|^2$. The result is
\be
{\cal N}^2
\int d^2z |z|^2 (\partial ^2 \frac{1}{{\bar z}^2}
+ c.c.) =  2{\cal N}^2  =4\pi \int \frac{d^2 z}{|z|^2} =
16\pi^2 log\epsilon + finite
\label{nine}
\ee
We do not  understand completely this construction; for example,
 if one integrates with another weight, the answer will be
 different;  Moreover,
 the  cut-off $\epsilon$ is a world-sheet
 cut-off, which is not the same
 as the one which will arise later due to divergences
 in the integral over $q$.

 These mathematical problems will disappear immediately,
 if we  postulate that  in the soliton  background
  $J$ are not the usual, but the {\bf logarithmic}  currents.
 Then, instead of (\ref{seven})
 one has to use (\ref{CC}) to obtain
\be
 <J^i(z)J^j(0)> = - \kappa  \delta ^{ij} \frac{\ln |z|^2}{z^2}
\label{correctnorm}
\ee
 which immediately leads to finite ${\cal N}^2$ and well-defined
Zamolodchikov metric
\bea
-\kappa {\cal N}^2
 |z|^4 \left(\partial^2 \frac{\ln |z|^2}{{\bar z}^2}
+ c.c.\right) =  2\kappa {\cal N}^2  = 4\pi; \nonumber \\
{\cal N}^2 = \frac{2\pi}{\kappa}
\label{correctN}
\eea
which implies that ${\cal N}^2$ is not logarithmically
divergent as in ~\cite{paban}, but finite.

How, then, can one reproduce the logarithmic divergences found
 in \cite{paban}, which are crucial for the recoil problem ?

 The answer was given at the end of the previous section -
 the same logarithmic operators lead to extra $\ln q$ terms.
 Let us explain this in more detail.
Given that the effects associated with a
change in the conformal field theory,
such as back-reaction of matter on the structure of space-time etc,
are purely stringy effects,
the most natural way of incorporating them in a $\sigma$-model
language is to go {\it beyond a fixed genus} $\sigma$-model
and consider the effects of resummation over world-sheet
genera. This is a very difficult procedure to be carried out
analytically. However,
for our purposes a sufficient analysis,
which describes the situation satisfactorily (at least at a
qualitative level), is that of a heavy extended object in target
space, which can be treated semi-classically. From a first-quantized
point of view,
this implies
resumming one-loop (torus)
world-sheets. Non-trivial effects arise from degenerate
Riemann surfaces, namely from long-thin world-sheet tubes (wormholes)
that are attached to a Riemann surface of lower genus (sphere
in this case).
>From a formal point of view, representing a Riemann surface
 $\Sigma$ as a tube  connecting
 two Riemann surfaces $\Sigma_1$ and $ \Sigma_2$ one can
take into account the degenerate
 tube by inserting a
complete set of intermediate string states
${\cal E}_\alpha$ ~\cite{polchinski}.
 Then the amplitude
\bea
\int dm_{\Sigma}
\langle \prod_{i} \int d^2\xi_i V_{i}(\xi_i)
\otimes (ghosts) \rangle
\eea
 is given by the following expression
\bea
  \sum _\alpha   \int dq d{\overline q}
  \int d^2 z_1 ~\int d^2 z_2  \int dm_{\Sigma_1\oplus\Sigma_2}
\nonumber \\
\langle \prod_{i} \int d^2\xi_i V_{i}(\xi_i)
 {\cal E}_\alpha (z_1)
\otimes  q^{L_{0} -1} {\overline q}^{{\overline
L}_{0}
-1}
\otimes
{\cal E}_\alpha (z_2) \otimes (ghosts)
\rangle_{\Sigma_1 \oplus \Sigma_2}
\label{props}
\eea
where $\int dm$ means integration over respective moduli space,
$V_{i}(\xi_{i})$ are the vertex operators for scattering states,
${\cal E}_\alpha$ are the complete set of the intermediate states
 with dimensions  $h_\alpha, {\overline h}_\alpha $
  propagating along the thin tube, connecting
the world-sheet pieces $\Sigma_1$ and $\Sigma_2$ (in the case
of the degenerating torus handle of interest to us,
$\Sigma_1 = \Sigma_2$).
The terms `ghosts' indicate
appropriate insertions of ghost fields.

One easily observes that
{\it extra} logarithmic divergencies in (\ref{props}),
may come from
states with $h_\alpha ={\overline h}_\alpha =0$~\cite{polchinski},
 in which case one has the infrared divergence  at small $q \rightarrow
0$  integral
$\int  dq d{\overline q}/ q {\overline q}$.
 A trivial example of such an operator is the identity,
which however {\it does} not lead to
non-trivial effects, since it carries
{\it zero} measure in the space of states, and hence
it does not contribute
to (\ref{props}).  On the other hand, if there are
states that are separated by a {\it gap} from the
continuum of states, i.e. {\it discrete} in the space of states,
then
they bear non-trivial contributions
to the sum-over-states and  lead to divergencies
in the amplitude (\ref{props}).

The logarithmic states we discussed earlier are precisely these states,
 and their  total dimension is zero because of the ghost insertion -
 actually they have  the form $\bar{c}c \otimes \partial _\alpha
J^\alpha$ \cite{paban} and ghosts shift the total anomalous dimension
 from $(1,1)$ to $(0,0)$.

 However we have to find another logarithmic divergence, instead
 of the normalization factor ${\cal N}^2$ - which is precisely the
 logarithmic terms in (\ref{CDprop}) arising due to the mixing between
 $C$ and $D$ states. After some algebra  we obtain expressions
 of the type
\bea
\frac{1}{\kappa} \int
\frac{ dq d{\overline q}}{ q {\overline q}} \left[ \ln q
\int d^2 z_1 D(z_1) \int d^2 z_2 C(z_2) +c.c.\right] \nonumber \\
 \sim
\frac{1}{\kappa} (\ln^2 \epsilon)
\int d^2 z_1 D(z_1) \int d^2 z_2 C(z_2) + c.c.
\eea
 which gives the leading singularity $\ln^2 \epsilon$. Besides this
 term there will be terms with $\ln \epsilon$ corresponding
 to $\int d^2 z_1 D(z_1) \int d^2 z_2 D(z_2)$ and
 $\int d^2 z_1 C(z_1) \int d^2 z_2 C(z_2)$ terms.
 The effects of a dilute gas of wormholes on the sphere exponentiate
 this bilocal operator and  one can obtain a change in the
world-sheet action (\ref{two})
\be
\Delta S \sim \frac{1}{\kappa} (\ln^2 \epsilon)
\int d^2 z_1 D(z_1) \int d^2 z_2 C(z_2)
\ee
This  bilocal term  can be writen as a local
world-sheet effective action
term, if one employs the well-known trick
of wormhole calculus~\cite{wormholes}
by writing
\be
  e^{\Delta S} \propto \int d\alpha_C d\alpha_D
  \exp\left[G^{mn}\alpha _m \alpha_n  + \alpha _C \frac{\ln\epsilon}{\kappa}
\int d^2 z C + \alpha _D \frac{\ln\epsilon}{\kappa}
\int d^2 z D \right]
\label{wlocal}
\ee
and, thus, it can be represented as a local deformation
on the world-sheet of the string but with a
coupling constant $\alpha _m$, where $m = C$ or $D$ and $G^{mn}$ is
 the metric necessary to reproduce the initial bilocal operator.

We shall not discuss  this problem  in more detail;
 further analysis  can be done along the lines of ref. \cite{paban},
 where   logarithmic divergences $\ln\epsilon$ in the case of
 elastic
 scattering on a soliton lead to the conservation of momentum.
 A more detailed picture  will be given in a future publication.
 Instead,
 in this article,
 we shall consider some exact CFT connected with string
 solitons and analyse some  conformal blocks in search of logarithmic
 operators.

\newsection{Logarithmic operators in exact CFT backgrounds}
\newsubsection{ CFT for string solitons}

To warm up and make it clearer what we mean, we start
by a simple four-dimensional effective action
involving antisymmetric tensor, dilaton and graviton fields.
The action reads
\be
S=-\frac{1}{\kappa ^2}\int d^4 x \sqrt{g}\{ R -
\frac{1}{2} (\partial \phi )^2
-\frac{1}{12} H_{\mu\nu\rho}^2 \}
\label{solaction}
\ee
A solution which could correspond to an extended object
in target space corresponds to the following
background configuration (5-brane)~\cite{solitons},
\bea
   g_{\mu\nu} &:&  ds^2 = e^{\phi(x)} dx^{\mu} dx_\mu
\nn \\
B_{\mu\nu} &:& H^{*}_{\mu} \equiv \epsilon^{\mu\nu\rho\sigma}
H_{\nu\rho\sigma} = \pm \partial _\mu e^{\phi(x)}
\label{solution}
\eea
where we wrote only the
dependence on 4 Cartesian coordinates $x^{\mu}$ in
a $4D$ transverse space.  The Bianchi identity for
$H^*$ , $\partial ^\mu H^{*}_\mu  =0$, implies
\be
   \nabla ^2 e^\phi = 0
\label{dilatoneq}
\ee
which leads to a  solution~\cite{solitons}
\be
e^\phi = e^{\phi _0} + \sum _{i=1}^n \frac{Q_i}{|x- a_i|^2}
\label{inst}
\ee
The $4D$ transverse space is asymptotically flat , at each singular
point $a_i$ a semiinfinite wormhole is glued in. To see this let us
restrict  ourselves to the simple case $n=1$ and neglect
$\exp(\phi_0)$ term. Then, one gets for the dilaton
\be
e^{\phi (x)} = Q/r^2
\label{sol2}
\ee
with $r^2 = \sum _{i=1}^4 x_i ^2$.
This leads to a linear dilaton~\cite{aben}
by a change of coordinates
\be
  \phi = -t/\sqrt{Q};
\qquad  t \equiv \sqrt{Q}
ln\sqrt{r^2/Q}
\label{lindil}
\ee
and to the metric
\be
 ds^2 = dt^2 + Q^2 d\Omega _3^2
\label{metric}
\ee
with $S^3$ a three sphere, and
$Q$ is  the axion charge defined as $Q = \int_{S_3} H$.
The geometry implied by (\ref{metric})
is that of a half throat consisting of a $S^3$ sphere of
radius $\sqrt{Q}$ times the open line $R^1$ and a linear dilaton
proportional to the coordinate on $R^1$, necessary for
conformal invariance due to the non-criticality of the target
dimension.
To this order in $\alpha ' \propto 1/k$
we can see, by looking at the dilaton $\beta$ function of the
theory, that the states of the theory are those of the tensor product
of a compact $SU(2)$ WZNW model of level $k=Q/\alpha ' \rightarrow \infty$,
$\alpha ' \rightarrow 0$, representing the sphere $S^3$ part and
the antisymmetric tensor, of central charge
\be
   c_{wzw} = \frac{3k}{k-2}
\label{wzw}
\ee
and a Feygin-Fuks representation for the
linear dilaton~\cite{aben}
and the time coordinate $t$ (half the real line),
with central charge
\be
   c_{FF}=1 + \frac{6}{k}
\label{ff}
\ee
The sum of the two central charges amounts to $4$ up to terms of order
$(1/k^2)$.
In the semi-classical  limit of  $k \rightarrow \infty$
($\alpha ' \rightarrow 0$), one can see that  the conformal blocks
of the WZNW theory \cite{kz} corresponds to the hypergeometric functions
of the form
\be
F(0, 0, 1; x),~~~F(1, 1, 2; x),~~~ F(0, 0, 0; x),
\label{hype}
\ee
 For the first one we definitely have logarithms  $\ln x$ in general
 solution (see section (2.1)), where for the last one
 the hypergeometric equation (\ref{hypereq}) becomes
\bea
x(1-x)\frac{d^2 {\cal F}}{dx^2} -x \frac{d{\cal F}}{dx} = 0
\eea
with the general solution ${\cal F} = A + B \ln(1-x)$,
 which has a logarithmic singularity at $x=1$.
 The situation
is tempting to suggest that there may be logarithmic operators
 in this case also, but the approximation of $k \rightarrow \infty$
is not really an exact result so as to justify the above argumentation.
\pr
The resolution comes
by including supersymmetry, as described in ref. \cite{solitons}.
In that case the four dimensions are pertaining to the internal
dimensions of some four dimensional
superstring constructions, but the details are irrelevant for
our purposes.
The important
point to notice is that
there are anomalies associated with left-right fermions,
 which effectively amount to modifying
the level $k$ of the supersymmetric WZNW model by $k -2$.
The resulting central charge of the WZW model is now
\be
     c_{WZW}=\frac{3(k-2)}{k} + \frac{3}{2}
\label{susy}
\ee
and that of the Feygin-Fuks /dilaton part
\be
 c_{FF} = 1 + \frac{6}{k} + \frac{1}{2}
\label{FFsusy}
\ee
where the factors of 3/2 in (\ref{susy}) and 1/2 in (\ref{FFsusy})
are due to the supersymmetric fermions of the CFT, which are
essentially free. The total central charge now is 6, and the
expansion in powers of $1/k$ {\it truncates}, leading to exact results
unlike the bosonic case.
\pr
Let us now locate the zero modes and the origin of logarithmic operators
in the superconformal blocks of the WZNW model.
 The conformal blocks were considered, for example,
in ref. \cite{fuchs}.
Consider the four point functions of a supersymmetric WZNW model
 with  the group $SU(N)$ - actually we need here only $N=2$.
The procedure is analogous to the
bosonic case~\cite{kz}, and one can derive easily the analogues
of the Knizhnik-Zamolodchikov equations that define the
superconformal blocks, $f_A^{(p)}$, $f_{\bar B}^{(q)}$
in a standard notation. In this particular case, the
solution acquires  the form
\be
f_{A{\bar B}}(x,{\bar x})=\sum _{p,q=1}^{2}a_{pq}f_A^{(p)}(x)
f_{{\bar B}}^{(q)}(\bar x)
\label{kz}
\ee
with some  constants $a_{pq}$ and
\bea
   f^{(1)}_1 (x) &=&x^{-2\Delta}(1 - x)^{{\tilde \Delta} - 2\Delta }
F\left(\frac{1}{k}, -\frac{1}{k}, 1-\frac{N}{k} ; x\right) \nn \\
   f^{(2)}_1 (x) &=&[x(1 - x)]^{{\tilde \Delta} - 2\Delta }
F\left(\frac{N+1}{k},\frac{N - 1}{k}, 1 + \frac{N}{k} ; x\right) \nn \\
   f^{(1)}_2 (x) &=&(k-N)^{-1}x^{1-2\Delta}
(1 - x)^{{\tilde \Delta} - 2\Delta }
F\left(1 + \frac{1}{k}, 1-\frac{1}{k}, 2-\frac{N}{k} ; x\right) \nn \\
   f^{(2)}_2 (x) &=&-N[x(1 - x)]^{{\tilde \Delta} - 2\Delta }
F\left(\frac{N+1}{k}, \frac{N-1}{k}, \frac{N}{k} ; x\right)
\label{fuchshy}
\eea
where
\be
\Delta = \frac{N^2-1}{2N k} ~~~~~
{\tilde \Delta} = \frac{N}{k}
\ee

It is also important to know that unitary highest weight
representations in this model exist only for non-negative
integer  $\hat{k} = k - N$ (see \cite{fuchs} and references therein),
 which means that minimal possible $k$ in this case is $N$, i.e.
minimal  $k=2$ in case of $SU(2)$.
  Taking into account
that in the model at hand the level $k$ is associated with
the size of the throat, i.e. the axion charge $Q$, we shall
 consider the ``minimal'' soliton.

In this case one can see immediately that  the dimension
of the adjoint field is ${\tilde \Delta} = 1$ and we have a degeneracy
 with all the consequences. Using (\ref{1+m}) and (\ref{1-m})
 one can see that at $k = N$ we have the  set
 of solutions similar to what was considered in  \cite{tsvelik}
\bea
   f^{(1)}_1 (x) &=&[x(1 - x)]^{1/N^2}
\ln x~F\left(\frac{N+1}{N},\frac{N - 1}{N}, 2; x\right) + H_2(x)
\nn \\
   f^{(2)}_1 (x) &=&[x(1 - x)]^{1/N^2}
F\left(\frac{N+1}{N},\frac{N - 1}{N}, 2; x\right)
\nn \\
   f^{(1)}_2 (x) &=&[x(1 - x)]^{1/N^2}
\ln x~F\left(1 + \frac{N+1}{N}, \frac{N-1}{N},  1; x\right) +H_1(x)
\nn \\
   f^{(2)}_2 (x) &=&[x(1 - x)]^{1/N^2}
F\left(\frac{N+1}{N}, \frac{N-1}{N},  1; x\right)
\label{logfuchs}
\eea
where $H_1$ and $H_2$ are two  functions discussed in
 section $(2.1)$ and
 we took  into account that  $1-2\Delta = 1/N^2$ for $k = N$.
Using the same methods as in \cite{tsvelik} we can  get the OPE
 for two primary fields in this model which has the form

\begin{equation}
V(z)~ V^{-1}(0) = |z|^{2/N^2 - 2}
\left\{ I +
z\bar{z}\left[{\cal D}(0) + {\cal C}(0) \ln|z|^2 \right] 
+ .....\right\}
\label{susyOPE}
\end{equation}

From this OPE one  can see immediately that the logarithmic
 pair ${\cal D}$, $ {\cal C}$
  has anomalous dimension $h = \bar{h} = 1$. These fields
 can be  constructed either  from 
the  adjoint field with $h = \tilde{\Delta} = 1$ and  
the  corresponding Kac-Moody currents, or  from  objects like
 $\bar{\partial}(C,D)$ and their complex conjugates. This situation is
 slightly different from the chiral situation of ref. \cite{tsvelik}. 
 At present we  are not in a position to determine the precise form
 of the  logarithmic
 pair.  Anyhow,   the above  analysis supports our conjecture that in the CFT
 corresponding to the  solitonic  background there
 are logarithmic operators with dimension $1$ which  generate
 the symmetry transformations.

It appears from the above  considerations
that logarithmic operators appear only for certain values
of the level $k$, namely $k =2$ for the model of $5$- brane.
 This is contrary to the generic arguments
given above on the existence of zero modes due to the
broken translational invariance by the center of mass of the solitonic
solution (instanton, monopole etc) , which appears to be a general
argument. The resolution is provided by the fact that so far we have
looked at part of the problem, only the supersymmetric  WZNW
model that refers
to the $S^3$ and the antisymmetric part of the dilaton.
The independence of $k$ may come  from combining (tensoring)
  this model with the Feygin-Fuks conformal field theory, and then
repeating the Kniznik-Zamolodchikov equation for the tensored
CFT. This hope is  based on the fact that only for $k=2$ we got the
 second operator with dimension 1. Let us note that in the case
of the CFT model
of disorder there were also two types of fields : a
 $SU(r)$ WZNW field and a $U(1)$ scalar field, and only by taking into
 account both of these fields one was able to recover
 the logarithmic operators
 in a general case. We hope the same is true in this
 model also.
 One, then, has to find the analogue of the hypergeometric functions
appearing in the conformal blocks of the tensor model in this case.
Once this is done , one should hopefully be able to show
the existence of logarithmic operators as a result of the
existence of a degenerate (zero) mode in the spectrum, other than
the identity, corresponding to `recoil' of the soliton center
of mass.
\pr

\newsubsection{$2D$ Black Holes}
\pr
As we discussed earlier there are  logarithmic operators
in the Liouville $c=1$ theory  which are
\bea
   C=\exp(\sqrt{2}\phi); \qquad D = \frac{1}{\sqrt{2}}\phi \exp(\sqrt{2}\phi)
\label{23}
\eea
It is interesting to know if this pair exists
 in a
two-dimensional black hole model which is a generalization
of the $c=1$ string theory. $2D$ black holes
 can be described  ~\cite{witt} as {\it exact}
conformal field theories of gauged WZNW type over the
coset $SL(2,R)/U(1)$ with  the central charge $c = 3k/(k-2) - 1$,
where $k$ is the level of the $SL(2,R)$ Kac-Moody algebra. At
$k = 9/4$ one has the critical string with $c=26$. One can see that
 the factor $1/(k-2) =4$ in this case, and for pure $SL(2,R)$ conformal
 blocks one again has integer $c$ in the
 corresponding hypergeometric
functions. However the conformal blocks for $2D$ black holes
correspond to the coset $SL(2,R)/U(1)$ and their  actual form is
yet  unknown.  So, instead of attacking
this problem
 in the  same way as before, we shall
 try to find some marginal deformations
 of the background which far from the horizon, where the space-time
corresponds to the $c=1$ Liouville theory, will have logarithmic
terms.  The $\sigma$-model action
describing a Euclidean black hole can be written
in the form
\be
S=\frac{k}{4\pi} \int d^2z \frac{1}{1+|w|^2}\partial _\mu {\bar w}
\partial ^\mu w + \dots
\label{threev}
\ee
where the conventional radial
and angular coordinates $(r,\theta)$ are given
by $w=sinh r e^{-i\theta}$ and the target
space $(r,\theta)$ line element is
\be
ds^2=\frac{dwd{\overline w}}{1 + w{\overline w}}=dr^2+tanh^2rd\theta^2
\label{fourv}
\ee
\pr
The corresponding exactly-marginal deformation,
which turns on matter backgrounds
in this geometry is
constructed by $W_\infty$ symmetry considerations~\cite{bakas,emn},
and
is given by \cite{chaudh}
\be
L_0^1{\overline L}_0^1 \propto
{\cal F}^{c-c}_{\frac{1}{2},0,0} + i(\psi^{++}-\psi^{--}) + \dots
\label{margintax}
\ee
where the $\psi$ denote higher-string-level operators \cite{chaudh},
and
the `tachyon' operator is given by
\be
{\cal F} ^{c-c}_{\frac{1}{2},0,0}(r)
=\frac{1}{coshr}
F(\frac{1}{2},\frac{1}{2}, 1; tanh^2r )
\label{tachyon}
\ee
There is an integral representation for the hypergeometric
 function \cite{ww}
\be
F(a, b, c; x) = \frac{\Gamma(c)}{\Gamma(b)\Gamma(c-b)}
\int_{0}^{1} t^{b-1} (1-t)^{c-b-1}(1-tx)^{-a} dt
\ee
which for
the hypergeometric function $F(1/2, 1/2, 1; x)$ gives us
\be
F(1/2, 1/2, 1; x) =
\int_{0}^{\pi/2} (1 - x \sin^2 \theta)^{-1/2} d\theta
\ee
and one sees immediately that at $x \rightarrow 1$ there
 is a logarithmic singularity. Actually one can show that
\be
F(1/2, 1/2, 1; x) = F(1/2, 1/2, 1; 1- x) \ln(1- x) + H(x)
\label{x1}
\ee
 where $H(x)$ is regular at $x \rightarrow 1$. Let us note
 that precisely this function appears as a conformal block in
 $c=-2$ model \cite{gurarie}.

The marginal deformation
(\ref{margintax}) contains the generalization
of the cosmological constant operator.
Asymptotically in space, $r \rightarrow \infty$, this operator
reduces to the flat-space $c=1$ exactly marginal deformation
of the  $c=1$ matrix model discussed above, which has
logarithmic operators. To see this let us consider
 $r \rightarrow \infty$, which means that $x = tanh^2r \rightarrow 1$.
>From  (\ref{x1})
 one observes that
\bea
{\cal F} ^{c-c}_{\frac{1}{2},0,0}(r)
=\frac{1}{coshr}
F(\frac{1}{2},\frac{1}{2} ; 1, tanh^2r ) = \nn \\
\frac{1}{coshr} \ln \frac{1}{cosh^2r}
F(\frac{1}{2},\frac{1}{2} ; 1, 1/cosh^2r ) +
\frac{1}{coshr}H(tanh^2r)
\nn \\
~~~ \nn \\
\rightarrow r\exp(-r) + A\exp(-r) + O(\exp(-2r)),
\eea
where $A$ is some constant. After appropriate change of
variables $r = \sqrt{2}\phi$ we see that at
$r \rightarrow \infty$ the background
${\cal F} ^{c-c}_{\frac{1}{2},0,0}(r)$ becomes
the $c=1$ cosmological constant
background including the logarithmic operator
$T(\phi) \sim (\phi -\frac{1}{2}ln~\mu)\sqrt{\mu}e^{-\sqrt{2}\phi}$.

\pr
Thus  the  exactly marginal deformation of  the black hole
 background also have logarithmic operators and one can hope
 that they can be observed  in the corresponding  conformal blocks.
 As wee saw, in the  case of black hole background
 the logarithmic  operators   are closely
 related to
 the hidden $W_\infty$ symmetry~\cite{emn}, however
 the full picture needs further clarifications.  It is
 an interesting problem to see how these new operators will
 affect the scattering amplitude in the black hole background.

\newsection{Conclusions}

 In this paper we have demonstrated that in the general class of
 solitonic string backgrounds
 a new phenomenon takes place - the emergence of
 logarithmic operators associated with a target space  symmetry.

 The physical meaning of the  target space  symmetry is
  the existence of the  normalizable zero modes corresponding to
 the deformations of the string background under the symmetry of the
full theory. The  simplest examples we considered support this picture
 and this is a pleasant fact.

 We think that the phenomenon we discussed is very general and
 it is important for studying the general problem of back-reaction
 in string theory - the recoil problem we considered is only
  one of the problems of this type. It will be interesting to
  expand   our methods to the case of general $P$- and $D$-branes.
 We hope to return to this,
 as well as to
other related problems, such as black holes in string theory etc,
 in the near future.

 It will be also interesting to  find a general classification
 of all possible logarithmic operators and to use this as a
world-sheet approach to classify the possible target space symmetries
 in  string theory.  At present this remains the
biggest challenge.

\newsection{Acknowledgments}
The authors  are grateful to J. Caux, J. Ellis,  G. Ross,
  M. Shifman  and especially to A. Lewis,  A.Tsvelik and J. Wheater for
  valuable and inspirational  discussions.
One of us (I.I.K) would like also to thank
 M. Flohr  and V. Gurarie for attracting our attention to
 papers \cite{flohr}, \cite{kausch}.

\newpage
{\renewcommand{\Large}{\normalsize}

\end{document}